\documentclass[intlimits,twoside,a4paper]{article}

\usepackage[cp1251]{inputenc}

\usepackage[eqsecnum]{cmpj3}

\usepackage{graphicx}
\usepackage{float}
\usepackage{changepage}
\usepackage{psfrag}

\newcommand{\angstrom}{\mbox{\,\AA}}

\newcommand{\ev}{\mbox{{\,eV}}}

\newcommand{\bc}{\begin{center}}
\newcommand{\ec}{\end{center}}

\issue{2025}{28}{4}{43802}
\doinumber{10.5488/CMP.28.43802}
\title[Be and Be-related impurities in diamond: DFT study]%
 {Be and Be-related impurities in diamond: density functional theory study}
\author[K. M. Etmimi, M. A. Ojalah, A. M. Abotruma]{K. M. Etmimi\orcid{0009-0008-7550-4509}\refaddr{label1}\thanks{Corresponding author: \email{k.etmimi@uot.edu.ly}.},
M. A. Ojalah\orcid{0009-0004-1216-9791}\refaddr{label2}, A. M. Abotruma\orcid{0009-0009-1901-0288}\refaddr{label3}}
\addresses{
\addr{label1} Physics department, Facaulty of scinece, University of Tripoli, Tripoli, Libya
\addr{label2} Physics department, faculty of Education, University of Tripoli, Tripoli, Libya
\addr{label3} High Institute for Comprehensive Professions (Al-Shmokh Institution), Tripoli, Libya
}
\Keywords{diamond, beryllium, nitrogen, n-type, p-type, first principle}

\date{Received April 1, 2024, in final form September 1, 2025}
\begin{document}

\maketitle
\begin{abstract}

First-principles density functional simulations were employed to 
investigate the geometries, electrical properties, and hyperfine 
structures of various beryllium-doped diamond configurations, including
 interstitial (Be$_i$), substitutional (Be$_s$), and beryllium-nitrogen
 (Be-N) complexes. The incorporation of Be into the diamond lattice
 is more favorable as a substitutional dopant than as an interstitial
 dopant, although both processes are endothermic. Interstitial Be could
 potentially exhibit motional averaging from planar to axial symmetry 
with an activation energy of 0.1 eV. The most stable Be$_s$ configuration
 has $T_{d}$ symmetry with a spin state of $S=1$. Co-doping with nitrogen
 reduces the formation energy of Be$_s$-N$_{n}$ $(n=1-4)$ complexes, which further decreases 
 as the number of nitrogen atoms increases. This is attributed
 to the smaller covalent radius of nitrogen compared to carbon,
 resulting in reduced lattice distortion. Be$_s$-N$_3$ and Be$_s$-N$_4$ 
co-doping introduces shallow donors, while Be$_s$ exhibits $n$-type 
semiconductivity, but the deep donor level renders it impractical
 for room-temperature applications. 
These findings provide valuable insights into the behavior of beryllium as
 a dopant in diamond and highlight the potential of beryllium-nitrogen 
co-doping for achieving $n$-type diamond semiconductors.
\printkeywords
\end{abstract}


\section{Introduction} \label{sec:intro}

Diamond is widely used in several fields due to its remarkable
 properties. Consequently, it is challenging to exploit diamond in 
electronic device technology, which is known to dope both natural and 
synthetic diamonds.
 Diamond with $p$-type conductivity has been observed for
B-doped diamond by ion implantation and gas-phase 
doping \cite{fontaine-APL-68-2264,prins-DRM-11-612,tshepe.70.245107,VOGEL-13-1822,TSUBOUCHI-14-1969,wu.71.113108,ueda-90-122102}. 
However, doping diamonds to obtain $n$-type semiconductors remains a
challenge.
 Nitrogen, as the dominant impurity in natural and synthetic diamonds, is
 a relatively deep donor ($1.7\ev$ below the conduction band minimum). 
Phosphorus is a promising candidate for an $n$-type dopant
with a donor level ($0.6\ev$ below the conduction band minimum~\cite{gheeraert-SSC-113-577}),
although with low ionization fractions at room temperature in addition to
low phosphorus solubility.  
Despite a reasonable concentration of P being 
reached~\cite{hasegawa-APL-79-3068}, intrinsic defects compensate the
 introduced
carriers, and many  P~atoms are not on the substitutional site, which can 
influence the number of carriers.

Sulfur has been reported to generate a shallow donor state in CVD 
diamonds~\cite{nakazawa-APL-82-2074,sakaguchi-PRB-60-R2139,gupta-APL-83-491},
although the energy levels remain controversial.
Prins~\cite{prins-PRB-61-7191} reported that oxygen incorporated by 
the ion-implantation technique yields shallow 
$n$-type diamond with an activation energy of approximately  0.32\ev.
 However, he suggested that the generation of electronic devices above
 $600^\circ$C deactivates the oxygen donors.
Czelej et.~al. \cite{czelej-MRS-1-1}  showed that doping diamond with 
arsenic and antimony
introduces shallow donor levels in the band gap at 0.5 eV below the 
conduction band minimum.
Most of the potential dopants highlighted previously possess either high
 activation energy or high formation energy, thereby impeding the 
effectiveness of doping and overall performance of diamond 
post-doping.

 Co-doping is a strategic approach involving the
 utilization of a minimum of two dopants, offering a viable solution to
 the issue of compensation.
Several researchers have attempted to achieve $n$-type diamonds using two 
types of dopants (co-doping), mainly  
B-O~\cite{liu-pnas-116-16}, B-N~\cite{hu-cec-19-4571}, 
Li-N~\cite{othman-DRM-44-1O}, and the effect of hydrogen on doped diamond 
\cite{sque-PRL-92-017402,dai-C-43-1009,goss-PRB-65-115207}. However, 
none of these materials yield $n$-type semiconductors.
Hu et. al.~\cite{hu-carbon-42-1501} showed that incorporating phosphorus with boron can improve the mobility and conductivity and lead to a higher consistency in the lattice structure.
Theoretical studies have shown that typical transition metal impurities such 
as titanium, vanadium, and chromium affect both the structural and electronic
 properties. The formation and transition energies of these impurities are
 crucial for understanding their impact on the semiconducting
 properties \cite{WANG2025130215,Wang-2025-865} of the host material. 

Beryllium-doped diamond was successfully obtained by microwave plasma-assisted chemical vapor deposition (MPCVD) \cite{ueda-bv-18-121}. 
Beryllium has also been introduced via
ion-implantation  \cite{ueda-drm-17-1269}, providing a further evidence 
for the possibility of incorporating Be into diamonds.
As there is no complete understanding of Be doping in diamonds,
there is a great motivation to pursue this issue further.

\section{Method}\label{sec:method}

First-principles calculations were conducted using the plane-wave
 pseudopotential software called AIMPRO~\cite{briddon-PSSB-217-131,rayson-CPC-178-128} within the framework of density
 functional theory (DFT). 
The structures were optimized using a conjugate gradient scheme until 
the total energy changed by less than $10^{-5}$ a.u.
In the simulation of the structures investigated in this research,
 216-atom simple-cubic supercells with a side length of $3a_0$ are typically utilized. This ensured that the Be atoms were
 closer to one another than any of the periodic images, except for the
 calculation of reorientation barriers using climbing nudged elastic
 bands, where 64 lattice sites were used for the calculation because of
the computational cost. 

The self-consistent 
modelling code utilized in this study relies on density-functional theory
 for modelling the candidate structures by employing a generalized gradient 
approximation~\cite{PhysRevLett.77.3865}. Sampling of the Brillouin zone
 was carried out using the Monkhorst-Pack scheme~\cite{monkhorst-prb-13-5188}
 using a uniform mesh of $2\times2\times2$ 
special $k$-points. The expansion of Kohn-Sham eigenvectors employed a
 Gaussia $n$-type orbital basis. The treatment of carbon and beryllium 
involved fixed linear combinations of $s$- and $p$-orbitals augmented
 with a set of $d$-functions to account for polarization, resulting in
 22 functions per atom. Nitrogen was treated using
 separate sets of $s$-, $p$-, and $d$-Gaussians of four different
 widths, yielding 40 functions per atom. The determination of the matrix 
elements of the Hamiltonian involved a plane-wave expansion of the
 density and Kohn-Sham potential with a cutoff of 300 a.u., leading to 
well-converged total energies with respect to the charge density basis.
 Diffusion barriers were determined by applying 
 the climbing nudged elastic band approach, as described in previous
 works \cite{henkelman-jcp-113-9901,henkelman-jcp-113-9978}.
Hyperfine interactions were obtained by reconstructing the all-electron
wave functions in the core region \cite{shaw-PRL-95-105502,blochl-PRB-50-17953}.
Electrical levels were obtained by reference to a marker 
systems~ \cite{goss-DRM-13-684}. The donor levels are
 obtained in 
comparison to $N_s^0$ with a level at $E_c-1.7\, \ev$ and the acceptor
 levels are obtained in comparison to $B_s^0$ with a level at $E_v+0.37\, \ev$.
The formation energy of $X$ in the charge-state $q$ was calculated
using~\cite{zhang-PRL-67-2339}

\begin{equation}
E^{f}(X,q)= E^{t}(X,q) - \sum_{i}{\mu_i} + q(E_v(X,q)+\mu_\text{e}) + \alpha_\text{M}\frac{q^2}{L\epsilon}.
\end{equation}
Here, $E^{f}(X,q)$ represents the formation energy of system $X$ in
a specified charge state $q$. $E^{t}(X,q)$ denotes the total energy of
system $X$ in the same charge state $q$, derived from computational
 methods, such as DFT.
 $\mu_i$ are the chemical potentials of the atomic species, with the sum over the atoms in the system, $E_v$ is the
 energy of the valence band top, $\mu_\text{e}$ is the electron  chemical potential  relative to the valence band top and $\alpha_\text{M}$ is a geometric term
(Madelung constant), $L$ is the length scale of the system (supercell
 dimension) and $\epsilon$ is the static dielectric constant of the medium.
 In this study, we investigated the chemical potential of Be from the 
hexagonal close-packed structure of pure Be bulk, whereas the chemical potential of nitrogen was based on NH$_3$.

\section{Results}\label{sec:results}

\subsection{Formation energy and solubility}
To investigate the complexity of the formation of 
the Be-doped diamonds, we computed the defect formation energy.

\begin{table}[!ht]
\begin{adjustwidth}{-0.1\textwidth}{-0.1\textwidth}
\caption{Formation energies of substitutional and interstitial Be-doped in diamond. $E_s^f(X)$ and $E_i^f(X)$ represent the formation energy of substitutional and interstitial Be, respectively
}
\label{table:FormationE-Be}
\bc
\center
  \begin{tabular}{l|l|l}
   \hline
  Structure& $E_s^f(X)$&$E_i^f(X)$   \\
    \hline
 Be-doped & $4.62 \ev$&$10.28 \ev$\\
   \hline
  \end{tabular}
\ec
\end{adjustwidth}
\end{table}

Table~\ref{table:FormationE-Be} lists formation energies of
 substitutional and interstitial defects in Be-doped diamond denoted
 $E_s^f(X)$ and $E_i^f(X)$, respectively. The formation energies of the 
two configurations are quantified at
 $4.62 \ev$ and $10.28 \ev$, respectively. This observation suggests
 that the incorporation of a beryllium atom into the diamond lattice is 
more favorable when occurring as a substitutional dopant. Importantly,
 the formation energy corresponding to $E_i(X)$ reaches a considerable
 value of $10.28 \ev$, which signifies that the process of introducing
 dopants into the diamond structure possesses significant challenges for a 
solitary Be atom. Such findings align with the conclusions drawn by Zhou
 et. al.  \cite{zhou-Springer}.
The reason is that the covalent bond radius of the Be atom $(0.96\angstrom)$ is
larger than that of C atom $(0.77\angstrom)$.
\begin{table}[!ht]
\begin{adjustwidth}{-0.1\textwidth}{-0.1\textwidth}
\caption{Defect formation energies of Be$_s$-N$_1$, Be$_s$-N$_2$, Be$_s$-N$_3$ and Be$_s$-N$_4$ 
complexes. E$^f(X)$ represents the defect formation energy.}
\label{table:FormationE-Be-N}
\bc
\center
  \begin{tabular}{l|l}
   \hline
    Structure&E$^f$(X)   \\
    \hline
Be$_s$-N$_1$&$+0.28 \ev$\\
Be$_s$-N$_2$&$-4.07 \ev$\\
Be$_s$-N$_3$&$-4.35 \ev$\\
Be$_s$-N$_4$&$-4.64\ev$\\
   \hline
  \end{tabular}
\ec
\end{adjustwidth}
\end{table}

On the other hand, we found that Be could be incorporated more easily
in the presence of nitrogen. The four structures (Be-N$_n$, $n=1,2,3$ and $4$)
 calculation results in table~\ref{table:FormationE-Be-N} 
clearly illustrate that the formation energies of the structures are
 low endothermic for single nitrogen and exothermic for the other.
The low formation energies of these impurities indicate that nitrogen and
 beryllium atoms exist as a whole in diamond instead of as individual dopants.
The covalent radius of nitrogen ($0.734\angstrom$), indicating a smaller
 lattice distortion, results in less internal strain, which contributes to the
construction of doped structures with low formation energy.

\subsection{The interstitial Be, Be$_i$\label{sec:B_i}}

\begin{figure}[!ht]
\psfrag{c1}{C$_1$}
\psfrag{c2}{C$_2$}
  \flushleft
  \begin{minipage}{0.30\textwidth}
    \flushleft (a)\\[0mm]
    \includegraphics[width=\textwidth,clip]{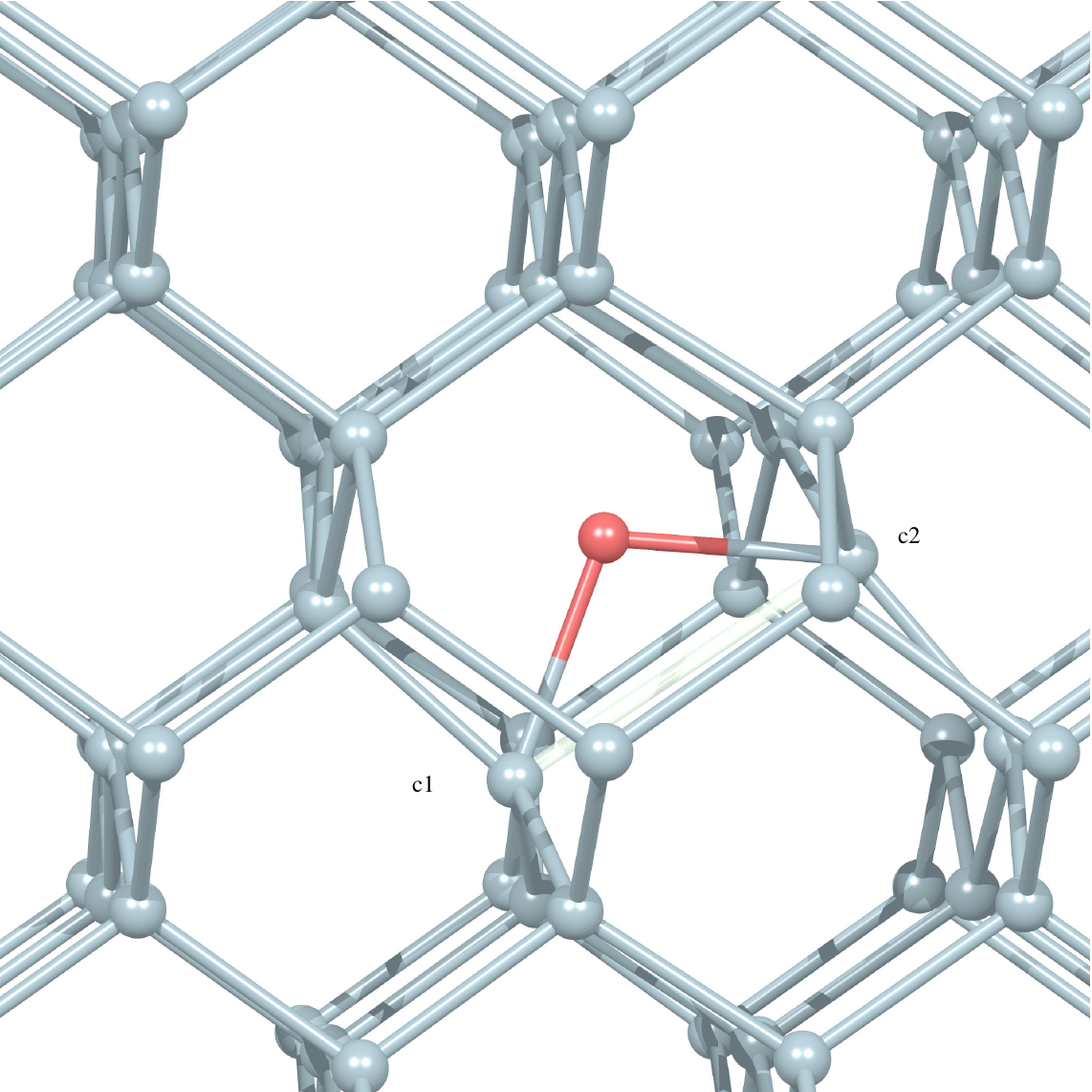}
  \end{minipage}
  \begin{minipage}{0.30\textwidth}
    \flushleft (b)\\[0mm]
    \includegraphics[width=\textwidth,clip]{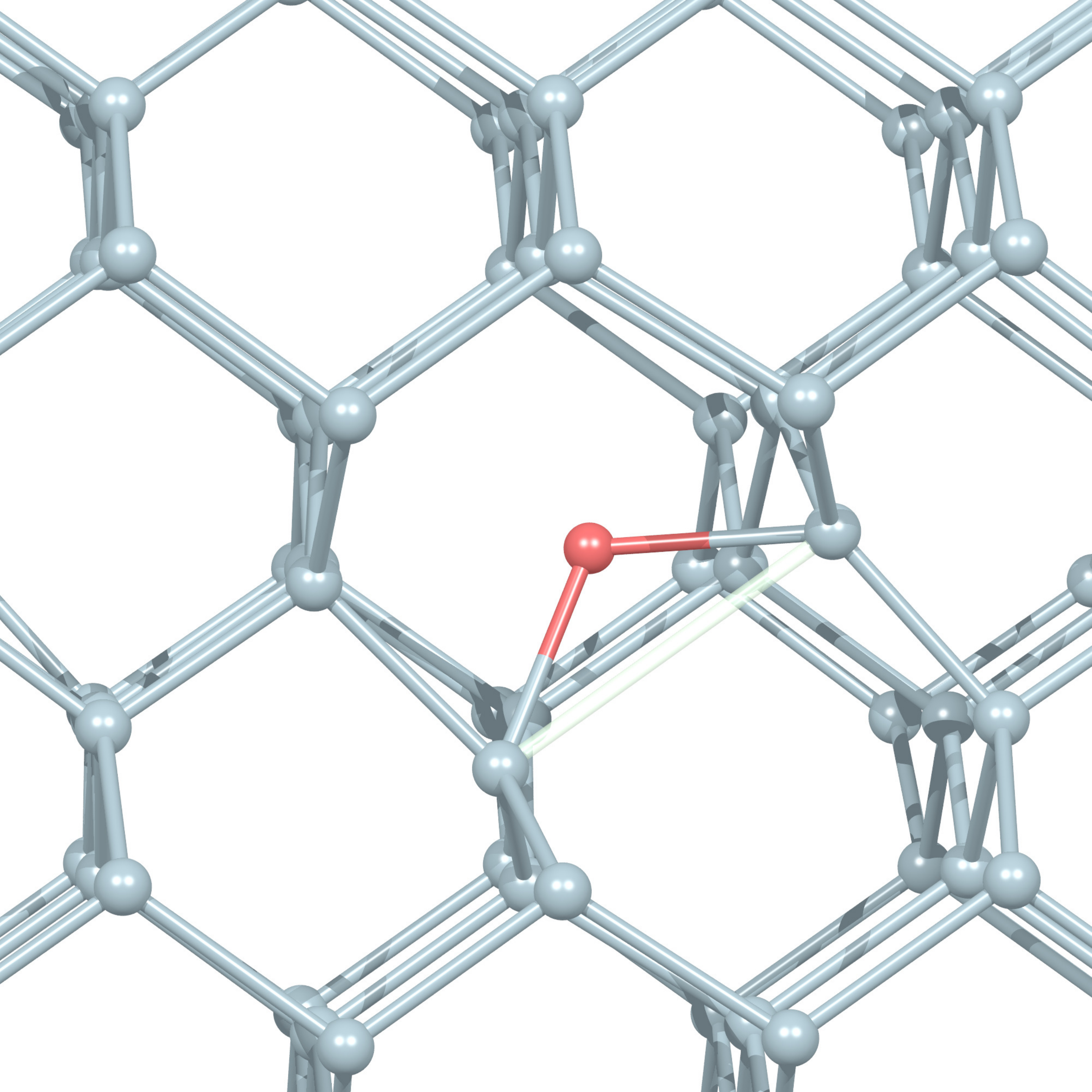}
  \end{minipage}
  \begin{minipage}{0.30\textwidth}
    \flushleft (c)\\[0mm]
    \includegraphics[width=\textwidth,clip]{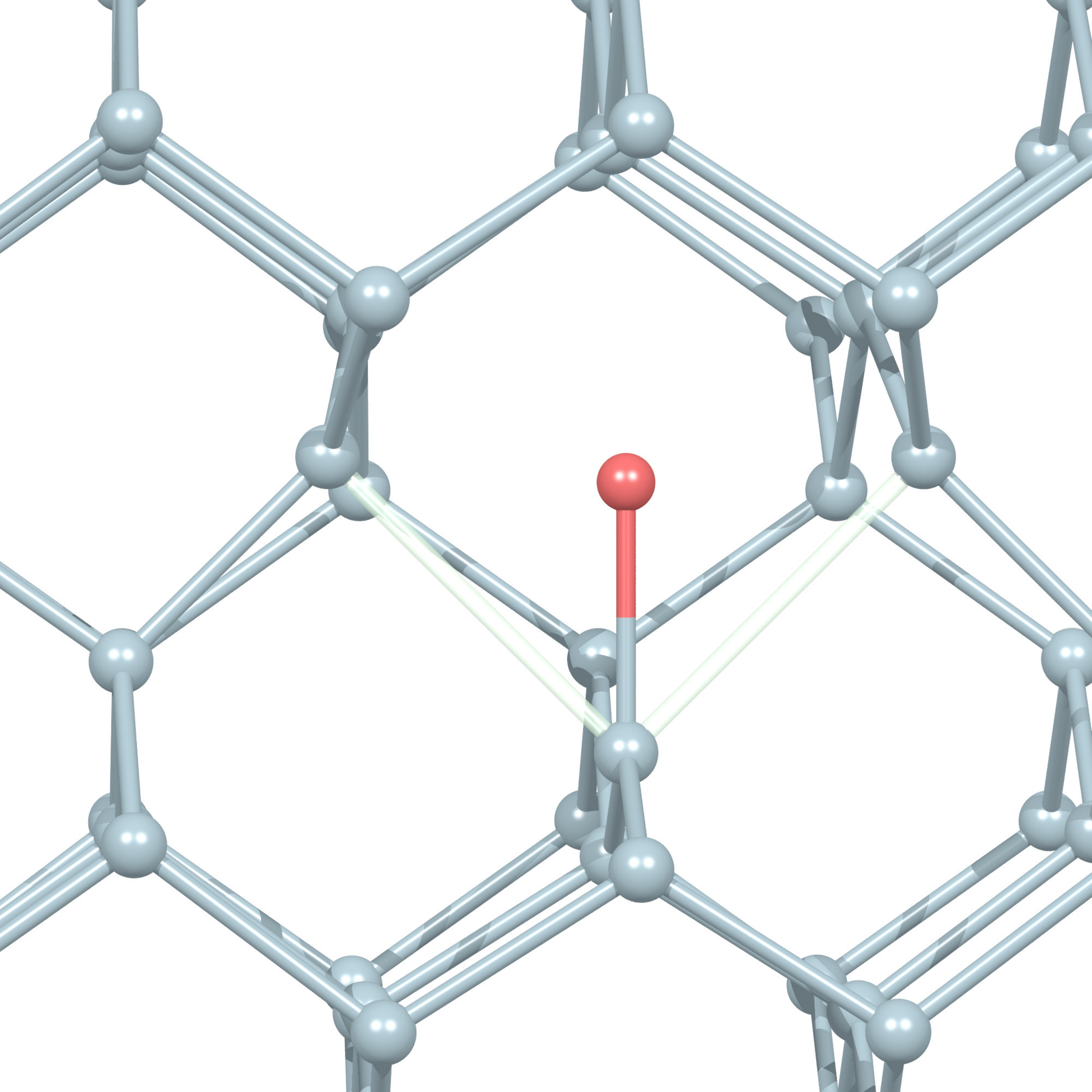}
  \end{minipage}
  \begin{minipage}{0.32\textwidth}
    \flushleft (d)\\[0mm]
    \includegraphics[width=\textwidth,clip]{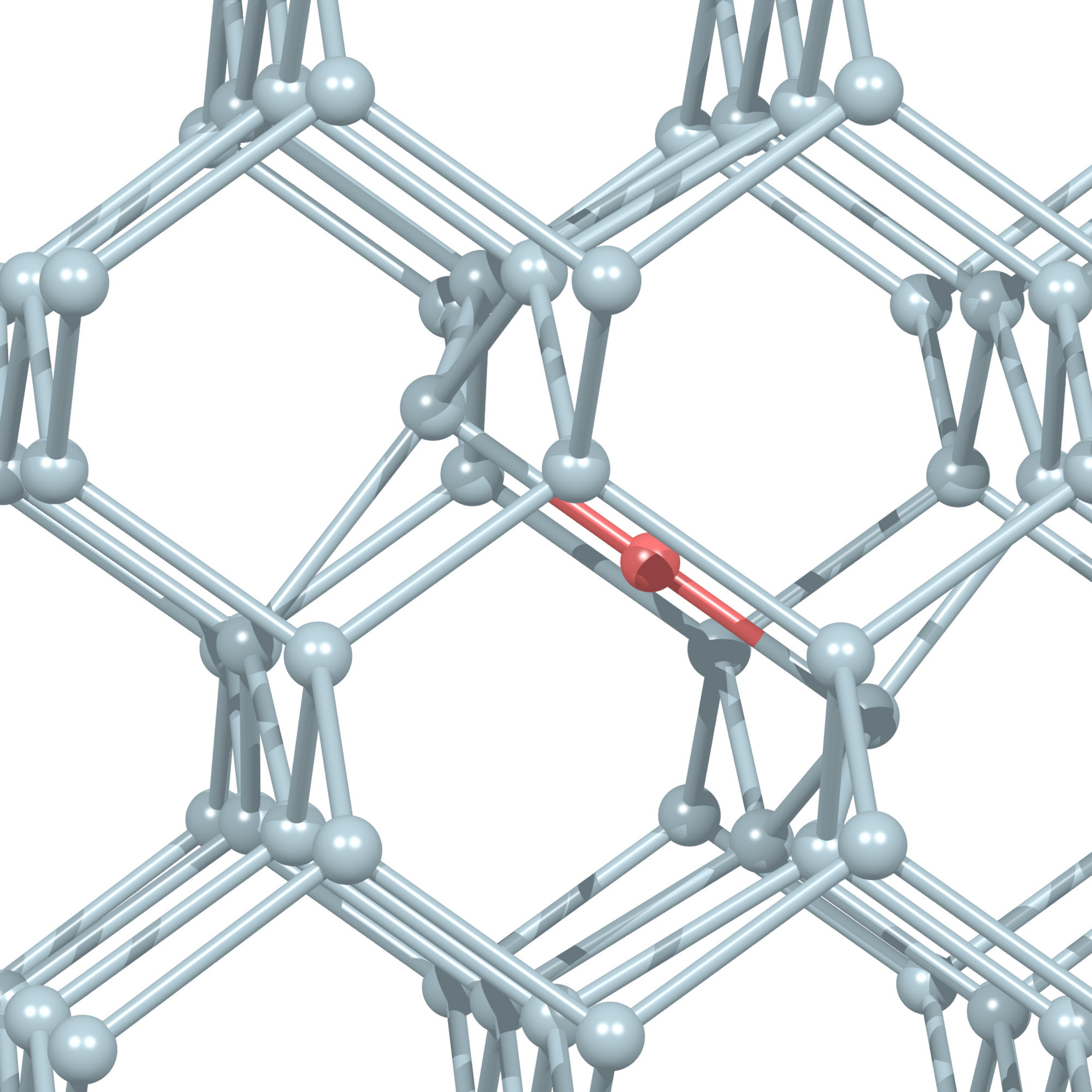}
  \end{minipage}
  \begin{minipage}{0.30\textwidth}
    \flushleft (e)\\[0mm]
    \includegraphics[width=\textwidth,clip]{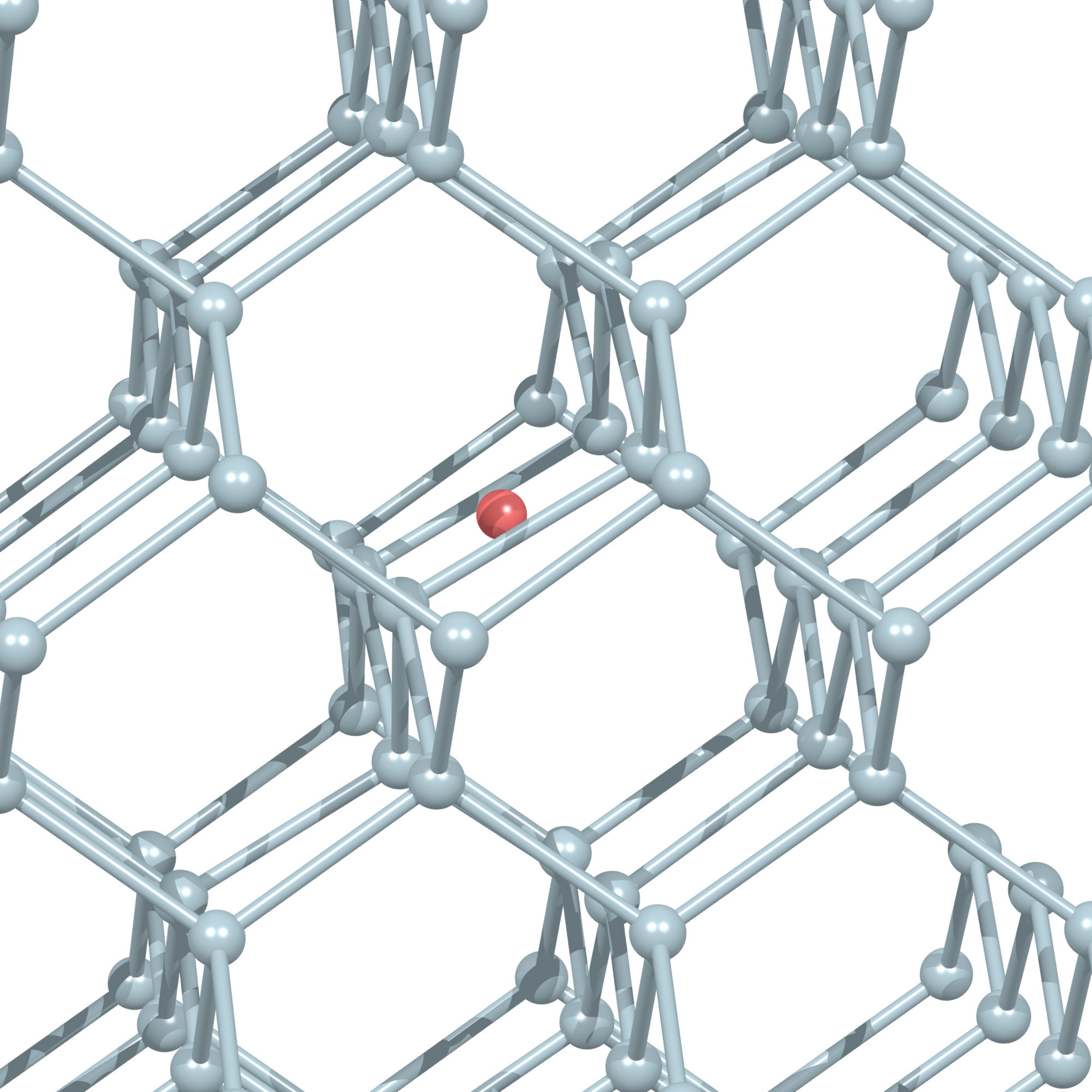}
  \end{minipage}
  \begin{minipage}{0.30\textwidth}
    \flushleft (f)\\[0mm]
    \includegraphics[width=\textwidth,clip]{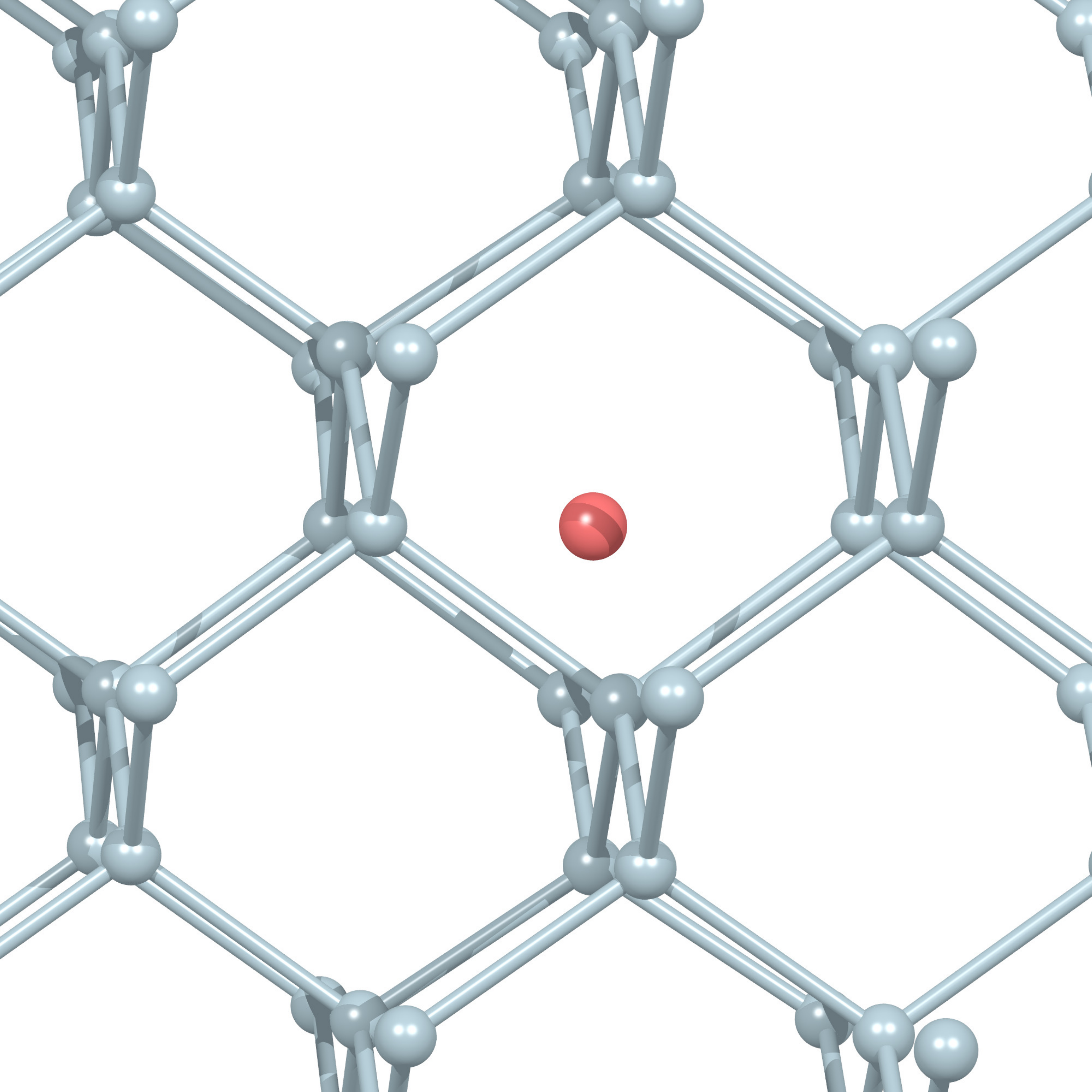}
  \end{minipage}
  \caption{(Colour online) Schematic structures of the substitutional Be configurations. 
Grey and red spheres represent C and Be, respectively (a) off-axis 
C$_{s}$, (b) off-axis C$_{2}$, (c) C$_{2v}$,  (d) BC-$D_{3d}$, (e) 
H-$D_{3d}$ and (f) T$_d$. Transparent sticks indicate broken bonds.}
   \label{fig:Bestructures}
\end{figure}

Beryllium atoms are placed at various interstitial positions in the 
lattice, including the bond center (BC), antibonding (AB), 
hexagonal (H), and 
tetrahedral (T$_d$) sites, as well as at random positions.
To establish convergence, all atoms in the cell were allowed to move 
freely during the geometry optimization.
Moreover, we varied the basis set and included or
excluded Be $1s$ electrons in the pseudopotential. However, no significant 
differences were observed.

\begin{figure}[!t]
\centering
  \begin{minipage}{0.6\textwidth}
    \includegraphics[width=1\textwidth,clip]{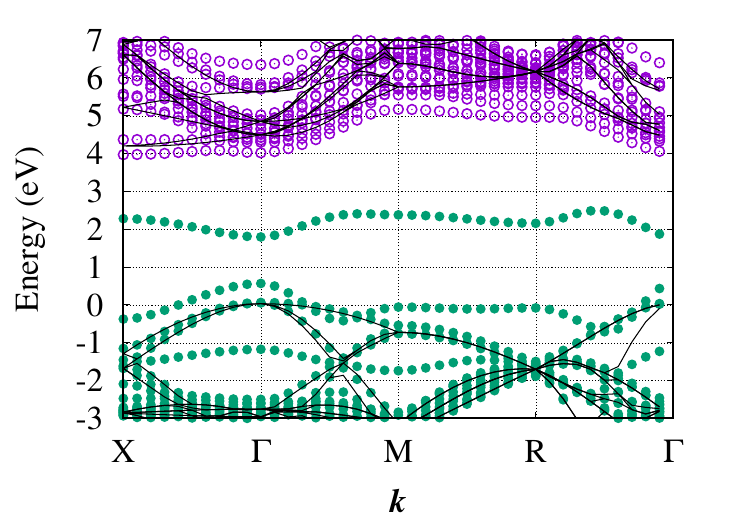}
  \end{minipage}
  \caption{(Colour online) The Kohn-Sham band structure in the vicinity of the band
 gap for Be$_i$  64 supercell.
Filled and empty circles show filled and empty bands, respectively,
with the bands from the defect-free cell superimposed in full lines
for comparison. The energy scale is defined by the valence band top
at zero energy ($E_v = 0 \ev$).
} 
   \label{fig:BC-Cs}
\end{figure}

In contrast to previous work~\cite{2009_Yan}, we found that among the six different structures studied, the
structure shown in figure~\ref{fig:Bestructures}(a) is the most stable.
The C-Be-C bond was heavily distorted from the [111] symmetry axis.
Since the atom is squeezed between two carbon atoms in normal lattice 
sites, the Be atom breaks a C-C bond and resides in an off-axis 
interstitial site in the [111] direction; consequently, the original
 $D_{3d}$ 
symmetry is reduced to $C_s$ symmetry during geometric relaxation
(denoted Be$_{i,oa1}$).
We found that the energy of $D_{3d}$ symmetry configuration was higher
than $C_s$ configuration by $1.54$\ev.
The Be atom has formed bonds with two nearest-neighbor C atoms,
with one Be-C bond being shorter than the others with lengths
of $1.44\angstrom$ and $1.49\angstrom$, respectively.
The broken C-C bond is dilated by approximately $52\%$ ($2.34\angstrom$).
The band structure (figure~\ref{fig:BC-Cs}) is suggestive of a donor
 level ($a^{\prime\prime}$ orbital) near the middle of the band gap 
($E_v+2.18 \ev$), 
which is fully occupied,
whereas the $a^\prime$ orbital is  almost resonant with the valence band
 edge.

The next most stable configuration is characterized by the beryllium (Be) atom in the off-axis interstitial arrangement (Be$_{i,oa2}$), as depicted in figure~\ref{fig:Bestructures}(b). In this configuration, the Be atom relocates to form equidistant bonds with two adjacent carbon atoms, each measuring $1.45\angstrom$. The resulting triangular structure, with the Be atom at the vertex, is displaced outside the $(110)$ plane, achieving C$_2$ symmetry. This configuration is less stable than Be$_{i,oc1}$ by $0.11\ev$.

\begin{figure}[!ht]
\centering
  \begin{minipage}{0.4\textwidth}
    \flushleft (a)\\[0mm]
    \includegraphics[width=1\textwidth,clip]{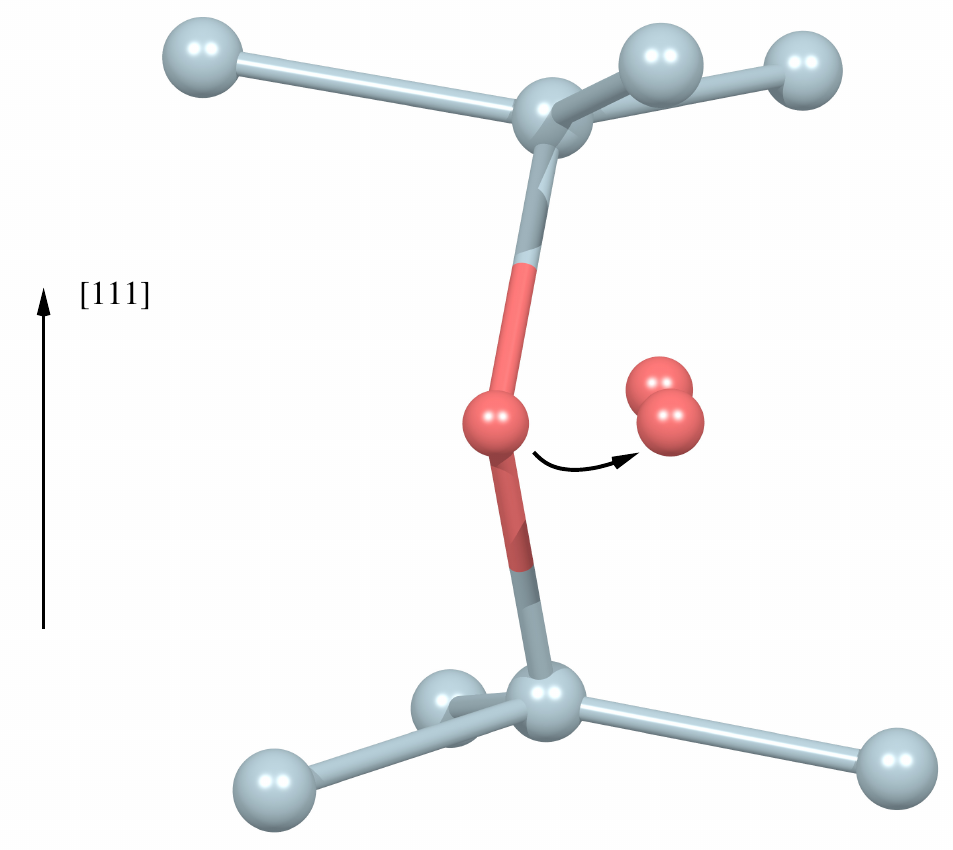}
  \end{minipage}
  \begin{minipage}{0.55\textwidth}
    \flushleft (b)\\[0mm]
    \includegraphics[width=1\textwidth,clip]{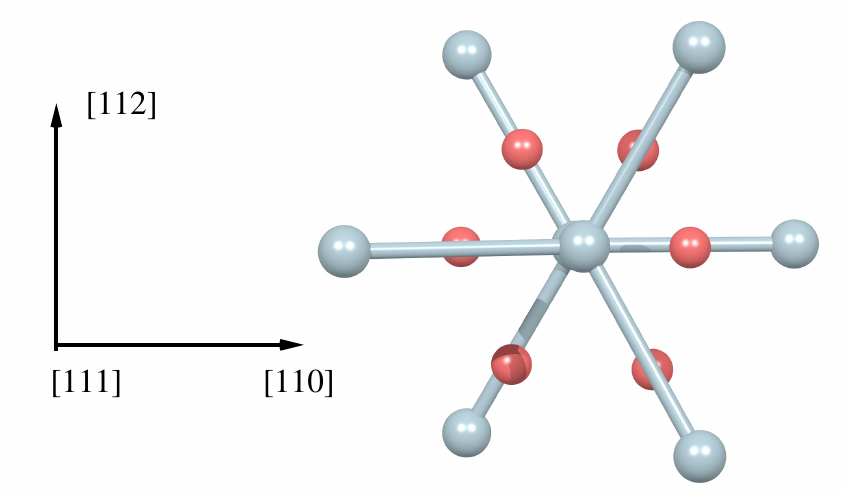}
  \end{minipage}
  \caption{(Colour online) Geometry structures of  the interstitial Be defects. (a)
 and (b) describes the bond centered configuration Be$_{i,oa1}$ with different orientation.}
   \label{fig:BC-Cs-C2}
\end{figure}

The results indicate a motional averaged structure for Be$_{i,oa1}$
defect as shown in figure~\ref{fig:BC-Cs-C2}(a), the Be atom orbits about
the symmetry axis. The barrier between the six equivalent distortions
 [figure~\ref{fig:BC-Cs-C2}(b)] from $D_{3d}$ symmetry was $0.11 \ev$. 
 The calculations suggest that at room temperature, there should be 
motional averaging from  $C_s$ symmetry to trigonal $D_{3d}$.

The structure shown in figure.~\ref{fig:Bestructures}(c) exhibited $C_{2v}$ 
symmetry. The Be atom forms a bond and resides at the mid-point 
between the three coordinated C atoms. This 
structure is higher in energy
than that of the $C_s$ configuration (Be$_{i,oa1}$) by $1.1 \ev$.

The structure in figure~\ref{fig:Bestructures}(d) shows that the Be atom
 sticks to the center of the C-C bond to undergo $D_{3d}$ symmetry.
The energy is higher than Be$_{i,oa1}$ by $1.54 \ev$.
In the structure where the Be atom forms a non-bond in a hexagonal 
configuration ($D_{3d}$ point group)
 [figure~\ref{fig:Bestructures}(e)], the energy is higher than that of
 Be$_{i,oa1}$ by $2 \ev$.
In contrast to the previous work \cite{2009_Yan}, we found that the
 structure in figure~\ref{fig:Bestructures}(f), which exhibits $T_d$ point
 group, is higher than that of Be$_{i,oa1}$ by $3.1 \ev$. The donor
 level was situated at $1.2\ev$ below the conduction band minimum. Yan~et. al. ~\cite{2009_Yan} suggested
that $T_d$ is an $n$-type metal conductivity characteristic which can readily be refuted.
The relative energies are listed in table~\ref{table:Bei}, which disagree
 with previous theoretical calculations \cite{Butorac-PRB-78-235204}.

\begin{table}[!ht]
\caption{Relative energy of relaxed structures with respect 
to $C_s$ structure in $\ev$.}
\label{table:Bei}
\bc
\center
  \begin{tabular}{c|c|c}
   \hline
    Symmetry&Spin configuration&Relative energy   \\
    \hline
$C_{2}$ &$S=0$&$0.11${\ev}\\
$C_{2v}$ &$S=0$&$1.11${\ev}\\
$D_{3d}$(BC) &$S=0$&$1.54${\ev}\\
$D_{3d}$(H) &$S=0$&$2.00${\ev}\\
$T_d$ &$S=0$&$3.14${\ev}\\
   \hline
  \end{tabular}
\ec
\end{table}

In the neutral charge state, the structure
is EPR-inactive. However, in the presence of shallow acceptors,
charge transfer is expected to occur, which gives rise to an EPR-active 
defect. Hyperfine tensors for the most stable configuration in the positive
 charge state are listed in table~\ref{table:Be_iHFI} and there are no 
hyperfine values for interstitial Be in the literature thus far.
In the positively charged state, the formation of the Be-C bond was
 weakened, with most of  the electron density localized at the
nearest carbon neighbors, leading to small anisotropic
hyperfine tensors for Be.
\begin{table}[!t]
  \caption{Calculated hyperfine tensors (MHz) of  Be and
    the two nearest neighbor $^{13}$C in interstitial Be, where $\theta$ 
is relative to  [001] and $\phi$ is measured from [100] toward [010] in
 xy-plane.}
\vspace{0.2cm}
\label{table:Be_iHFI}
\center
  \begin{tabular}{lrlrlrl}
 \hline \hline
Species
    & \multicolumn{2}{c}{$A_1$}
    & \multicolumn{2}{c}{$A_2$}
    & \multicolumn{2}{c}{$A_3$}\\    \hline
\multicolumn{7}{c}{$C_{s}$, charge$=+1$, $S=1/2$}\\
Be           &$5$   &$(90,315)$ &$1.2$    &$(165,45)$  &$-1.6$  &$(75,45)$\\
C$_1$            &$238$ &$(56,45)$ &$113$  &$(146,45)$ &$112$ &$(90,315)$\\
C$_2$           &$254$   &$(56,45)$ &$102$    &$(146,45)$  &$101$   &$(90,315)$\\

\hline
 \hline \hline
  \end{tabular}
\end{table}

\subsection{The substitutional Be, Be$_s$\label{sec:Bes}}
We also examined the stability and electronic characteristics of the 
substitutional dopants Be$_s$, 
because the interstitial lithium within the diamond lattice may exhibit
 mobility and is likely to be localized at carbon vacancies~\cite{job-DRM-5-757,prawer-APL-63-2502}. As the elements are
 located in the same period as lithium, it is conceivable that beryllium 
interstitial impurities can also be accommodated at the vacancy sites.
\begin{table}[!b]
	\caption{Calculated hyperfine tensors (MHz) of  Be and
		the four nearest neighbor $^{13}$C in substitutional Be. 
		$\theta$ and $\phi$ are given as indicated in table~\ref{table:Be_iHFI}}
	\vspace{0.3cm}
	\label{table:Be_sHFI}
	\center
	\begin{tabular}{lrlrlrl}
		\hline \hline
		Species
		& \multicolumn{2}{c}{$A_1$}
		& \multicolumn{2}{c}{$A_2$}
		& \multicolumn{2}{c}{$A_3$}\\    \hline
		\multicolumn{7}{c}{$T_{d}$, $S=1$}\\
		Be           &$-4$   &$(90,0)$ &$-4$    &$(90,90)$  &$-4$  &$(0,0)$\\
		C$_1$,C$_2$,C$_3$,C$_4$            &$57$ &$(55,225)$ &$17.2$  &$(45, 90)$ &$17.2$ &$(66, 333)$\\
		
		\hline
		\hline 
	\end{tabular}
\end{table}

\begin{figure}[H]
\psfrag{c1}{C$_1$}
\psfrag{c2}{C$_2$}
\psfrag{c3}{C$_3$}
\psfrag{c4}{C$_4$}
\psfrag{up}{Up}
\psfrag{dn}{Dn}
\centering
  \begin{minipage}{0.30\textwidth}
    \flushleft (a)\\[0mm]
    \includegraphics[width=1\textwidth,clip]{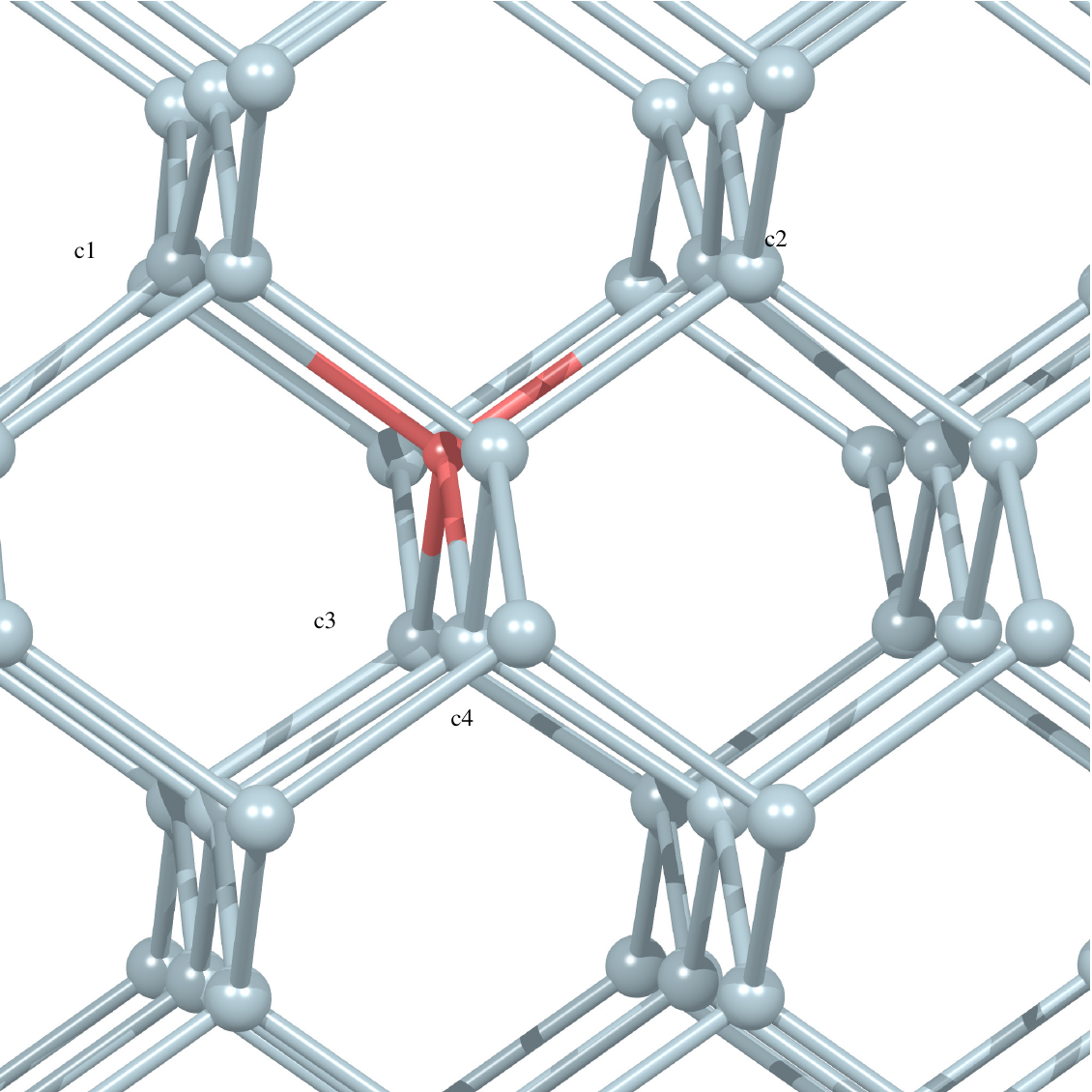}
  \end{minipage}
  \begin{minipage}{0.44\textwidth}
    \flushleft (b)\\[0mm]
    \includegraphics[width=1\textwidth,clip]{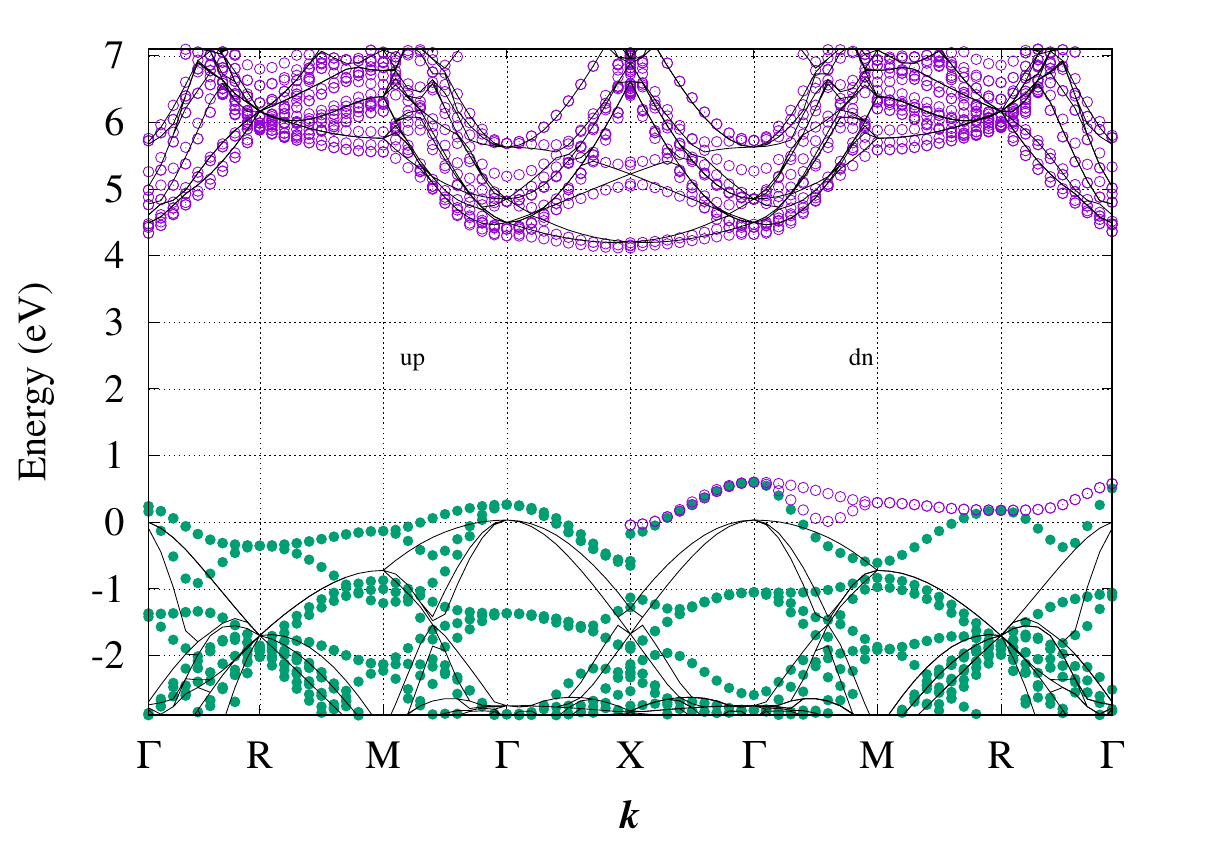}
  \end{minipage}
  \caption{(Colour online) (a) Schematics for Be$_s$ ($T_d$). Grey and red spheres represent C and Be, respectively. (b) Band structure for Be$_s$ configuration. Filled and empty circles show filled and empty bands, respectively. The energy scale is defined
by the valence band top at zero energy.}
   \label{fig:Bes-substu}
\end{figure}

Series of structures are determined. In contrast to Yan et. al. ~\cite{2009_Yan},
 we find that Be$_s$ has $T_{d}$ symmetry and is the most stable one
[figure~\ref{fig:Bes-substu} (a)]
with the spin state of the system being 1, the energy difference between the optimized
 structures with lowest energy, and that constrained to ($C_{3v}, S=0$), 
($C_{2v}, S=0$) and ($T_{d}, S=0$) are found to be marginally higher
 by $0.09\,\ev$, $0.10\,\ev$ and $0.12\,\ev$, respectively. 
The Be$_s$-C bond length was $1.67\angstrom$. The band-gap is 
shown in figure~\ref{fig:Bes-substu}(b).
The empty bands being $0.71\,\ev$ above the valence band top suggests that Be 
has double acceptor. 
A previous study~\cite{2009_Yan} showed that Be$_s$ induces a shallow acceptor 
level in the band gap. However, they reported neither a
 method that was used nor the location of the electrical levels.
For $S=0$ configuration, the acceptor level is at $0.60\ev$ above the 
valence band top.
Tests utilizing 216-atom cells produce qualitatively analogous outcomes
 for shallow acceptor defects; however, the energy levels are 
marginally lower within the band gap. 
This defect has three metastable structures, all of which possess
 acceptor levels located at $0.71\,\ev$ above the valence band.

In the neutral charge state, the structure
is EPR-inactive. However, the acceptor level in the lower half 
of the 
bandgap is likely to be negatively charged in materials
containing nitrogen-donors.
The small hyperfine tensors on the Be atom indicate that the  probability
 density of the two unpaired electrons is localized on the nearest-neighbor carbon atoms, as illustrated in table~\ref{table:Be_sHFI}.

\subsection{The Be$_s$-N$_n$ complexes ($n=1,2,3$ and $4$)}

We attempted to co-dope Be with nitrogen impurities.
Four sets containing one, two, three, and four nitrogen atoms were introduced
 around the substitutional Be atom to replace the original nearest-neighbor carbon atoms denoted Be$_s$-N$_1$, Be$_s$-N$_2$, Be$_s$-N$_3$ and Be$_s$-N$_4$, respectively.

\begin{figure}[!t]
	\psfrag{be}{Be}
	\psfrag{n1}{N$_1$}
	\psfrag{n2}{N$_2$}
	\psfrag{n3}{N$_3$}
	\psfrag{n4}{N$_4$}
	\psfrag{c}{C}
	\psfrag{c1}{C$_1$}
	\psfrag{c2}{C$_2$}
	\psfrag{c3}{C$_3$}
	\psfrag{n}{N}
	\centering
	\begin{minipage}{0.35\textwidth}
		\flushleft (a)\\[0mm]
		\includegraphics[width=1\textwidth,clip]{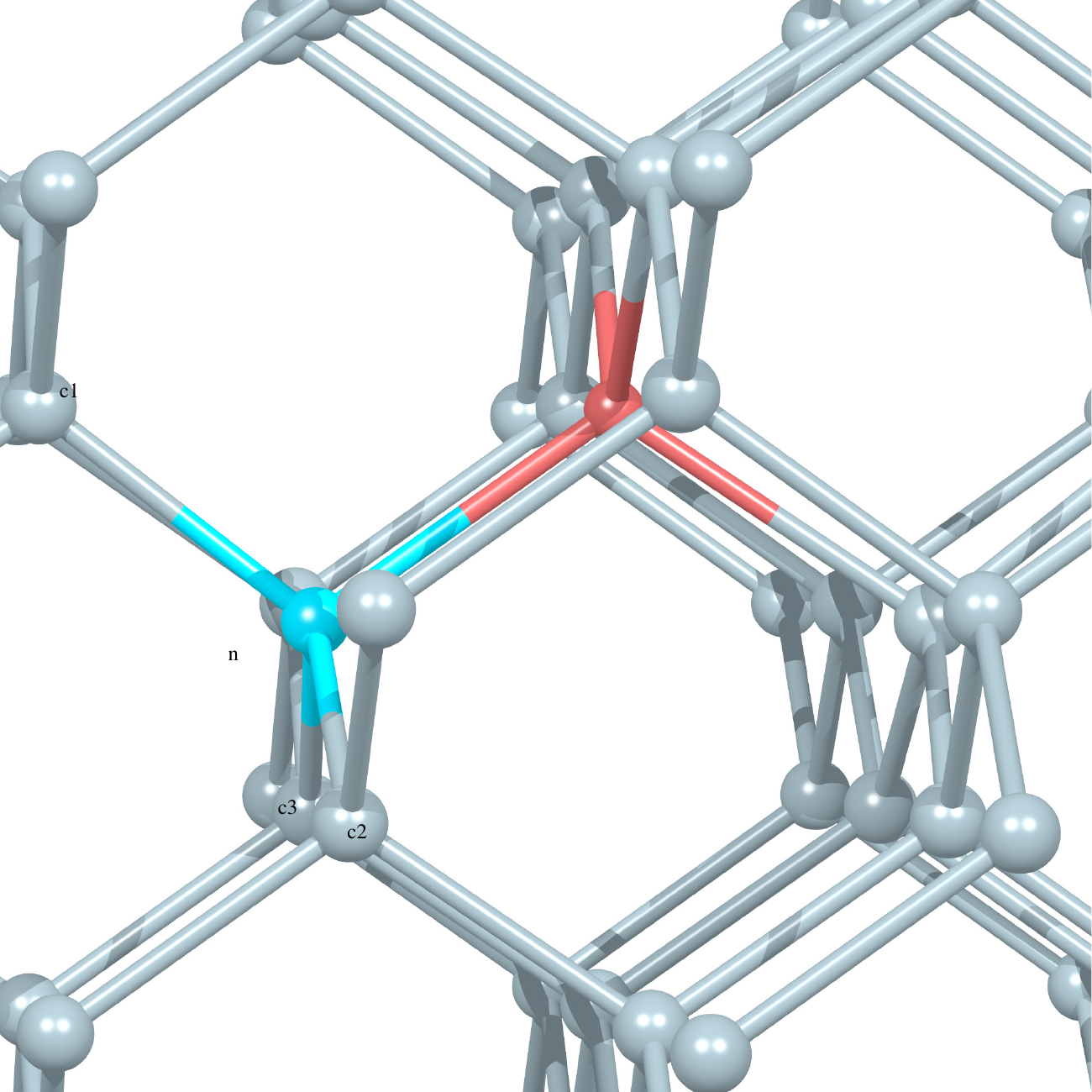}
	\end{minipage}
	\begin{minipage}{0.35\textwidth}
		\flushright (b)\\[0mm]
		\includegraphics[width=1\textwidth,clip]{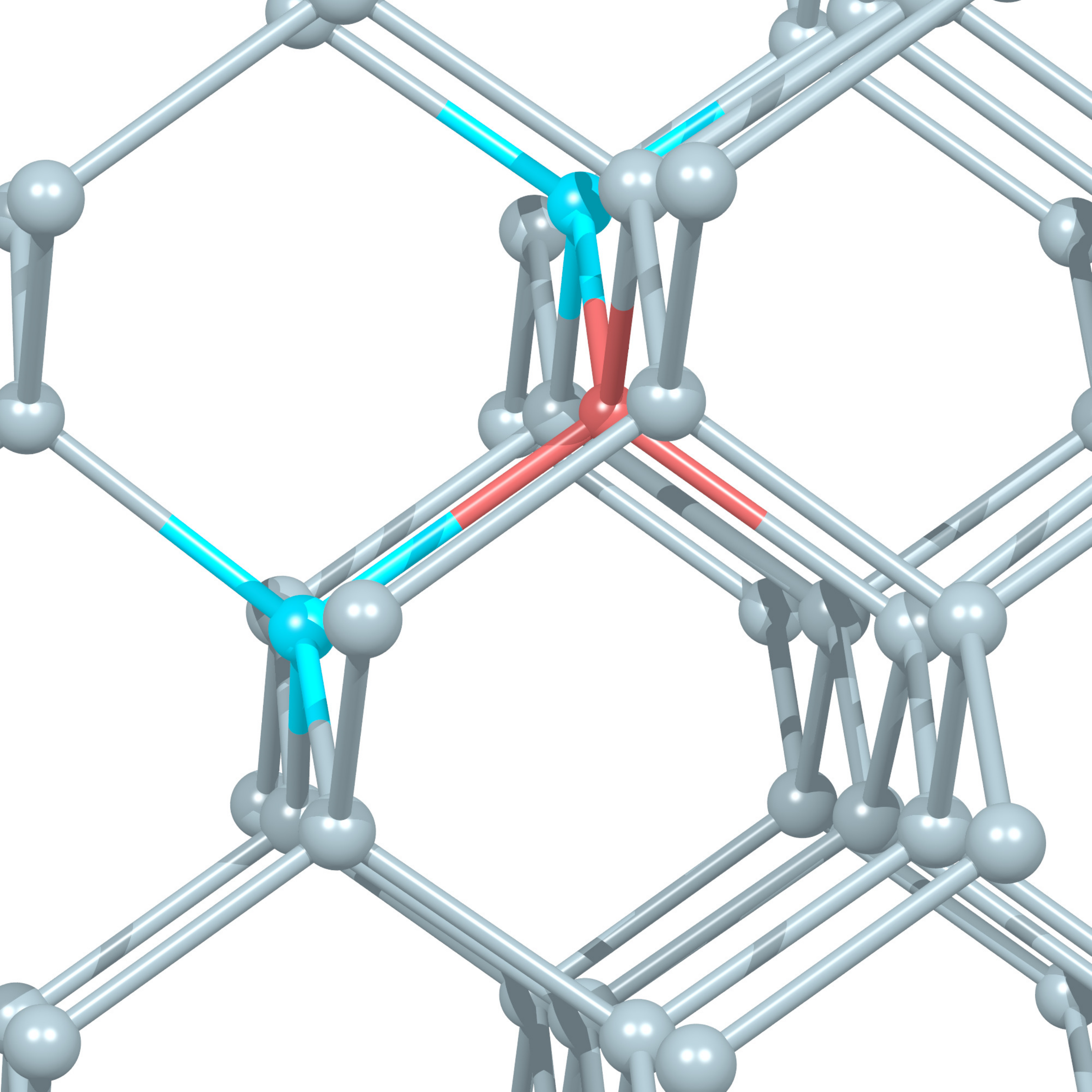}
	\end{minipage}
	\begin{minipage}{0.35\textwidth}
		\flushleft (c)\\[0mm]
		\includegraphics[width=1\textwidth,clip]{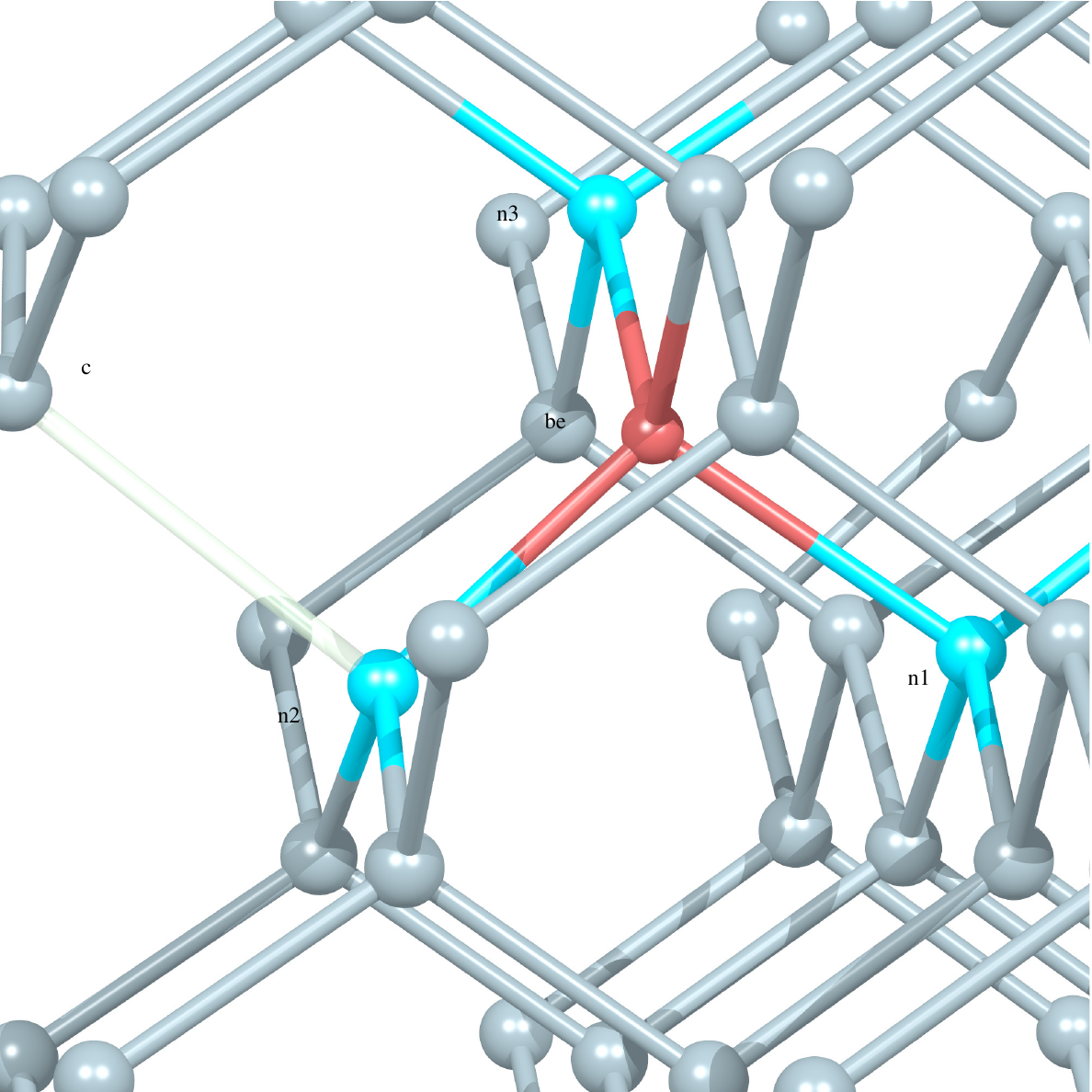}
	\end{minipage}
	\begin{minipage}{0.35\textwidth}
		\flushright (d)\\[0mm]
		\includegraphics[width=1\textwidth,clip]{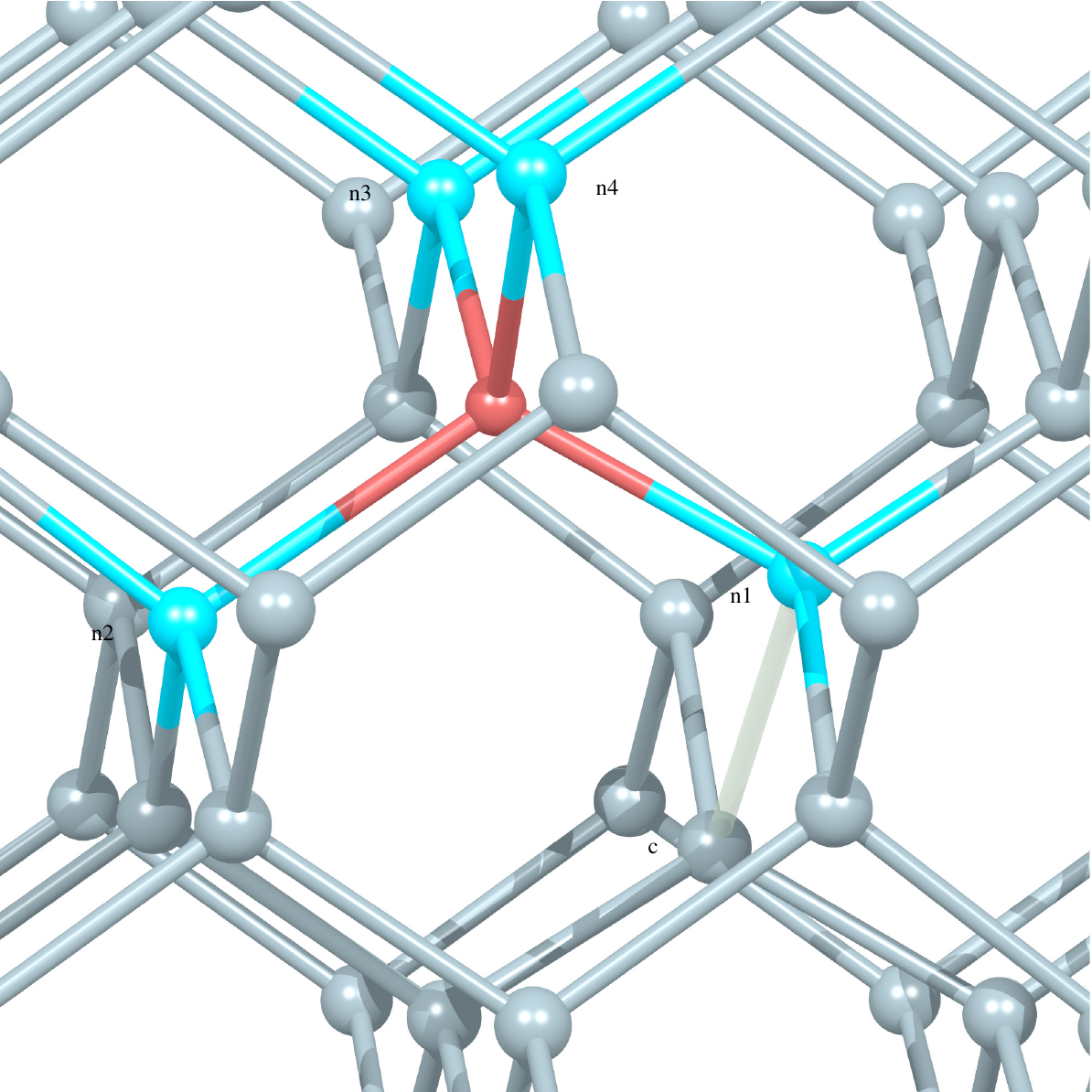}
	\end{minipage}
	\caption{Schematic structures of Be-N complexes.
		Grey, blue and red spheres represent C, N and Be, respectively.
		(a) Be$_s$-N$_1$, $C_{3v}$ symmetry and (b) Be$_s$-N$_2$, $C_{2v}$
		symmetry,  (c) Be$_s$-N$_3$, $C_{1}$ symmetry and (d) Be$_s$-N$_4$,
		$C_{s}$ symmetry. Transparent sticks indicate broken bonds.
	}
	\label{fig:Bes-3N}
\end{figure}

Be$_s$-N$_1$ relaxed into a  $C_{3v}$ symmetry as shown in
 figure~\ref{fig:Bes-3N}(a), where the Be-C and Be-N bond lengths are 
$1.67\angstrom$ and $1.60\angstrom$, which are respectively $8.4\%$ and
 $3.9\%$ 
longer than the C-C bonding. Our calculations show that the formation
 energy of Be$_s$-N$_1$ decreases ($+0.28\ev$)
compared to single-element doping at substitutional or 
interstitial sites. One possible explanation is that the atomic
radius of the nitrogen atoms is smaller.
Be$_s$-N$_1$ doping introduces an acceptor level lying at $0.85\ev$ above the
 valence band top. In the neutral charge state, the (Be$_s$-N$_1$) complex
 has an odd number of electrons, rendering it EPR active with $S=1/2$. 
The calculation hyperfine tensors for the 
N, Be and  the three carbon atoms nearest neighbor to Be are listed in table~\ref{table:Be_s-1N-HFI}.
The unpaired electron probability density is localized on the three carbon
 atoms neighboring the Be atom.
\begin{table}[H]
  \caption{Calculated hyperfine tensors (MHz) of  Be$_s$-N$_1$. 
$\theta$ and $\phi$ are given as indicated in table~\ref{table:Be_iHFI}}
\label{table:Be_s-1N-HFI}
\center
  \begin{tabular}{lrlrlrl}
 \hline \hline
Species
    & \multicolumn{2}{c}{$A_1$}
    & \multicolumn{2}{c}{$A_2$}
    & \multicolumn{2}{c}{$A_3$}\\    \hline
\multicolumn{7}{c}{$C_{3v}$, $S=1/2$}\\
Be           &$-0.2$   &$(125.3,135)$ &$-5.8$    &$(53.3,135)$  &$-5.8$  &$(90,45)$\\
N           &$-1.6$   &$(125.3,135)$ &$-1.2$    &$(144.7,-45)$  &$-1.2$  &$(90,45)$\\
C$_1$,C$_2$,C$_3$            &$27.6$ &$(116.8,120.3)$ &$29.3$  &$(45,180)$ &$84.2$ &$(57,49.5)$\\

\hline
 \hline 
  \end{tabular}
\end{table}

Interestingly, when the number of N atoms increased to two, the formation 
energy of Be$_s$-N$_2$ became negative ($-4.07$\ev). Increasing the 
number of N atoms results in less strain, as the 
covelent radius of the nitrogen atom is smaller than that of carbon.  Be$_s$-N$_2$ relaxed to $C_{2v}$ symmetry, as shown in
 figure~\ref{fig:Bes-3N}(b).
The Be-C and Be-N bond lengths are $1.63\angstrom$ and
$1.64\angstrom$, which are $5.8\%$ and $6.5\%$ longer than 
 the C-C bonds, respectively.
The ground state of the neutral complex would have $S=0$, and thus would
be EPR-inactive. 
Moreover, it possesses a very deep acceptor at $E_v+4.91\ev$. Therefore, it
is not possible for the complex to be ionized. 

When the number of N atom increases to three, the formation energy of 
Be$_s$-N$_3$ impurity is even lower ($-4.35$\ev), and it is $8.97\ev$ and
 $4.63\ev$ lower than that of Be$_s$-N$_1$ and Be, respectively.
Surprisingly, in contrast to Sun study~\cite{Yang-thesis},  
geometry optimization of Be$_s$-N$_3$  
resulted in a structure with C$_{1}$ symmetry. As shown in the 
figure~\ref{fig:Bes-3N}(c), the distances from Be to the two 
four-fold coordinated N atoms, three-fold coordinated N atom and  C are
 $1.65\angstrom$ $1.56\angstrom$ and $1.63\angstrom$, respectively, which are
 $7.1\%$, $1.3\%$ and 
$5.8\%$ longer than the original C-C bond.

\begin{figure}[!t]
\psfrag{n}{N}
\psfrag{Be}{Be}
\centering
  \begin{minipage}{0.35\textwidth}
    \flushleft (b)\\[0mm]
    \includegraphics[width=1\textwidth,clip]{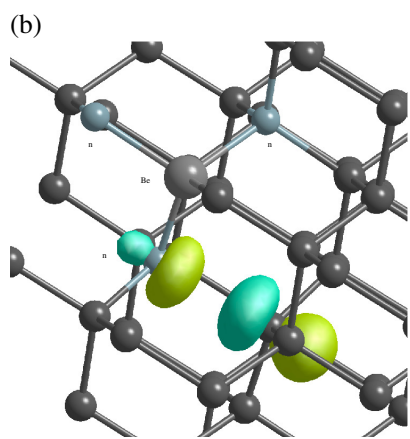}
  \end{minipage}
  \caption{(Colour online) Unpaired electron Kohn-Sham functions for Be-N$_3$ complex.}
   \label{fig:isos-Be-3N}
\end{figure}

Be$_s$-N$_3$ is also found to possess a metastable structure with C$_s$
symmetry; this configuration is only  $+0.11 \ev$ higher
 in energy than the $C_1$ configuration.
in contrast to the results of a previous study~\cite{Sun-iop-2024,Yang-thesis}. The band gap indicates a shallow  donor level at $1\ev$ below the
 conduction band minimum which is lower than that of N.
The hyperfine tensors are presented in table~\ref{table:Be_s-3N-HFI}.
The unpaired electron density was largely localized on C-N antibonding
 orbital as illustrated in figure~\ref{fig:isos-Be-3N}, providing a hyperfine tensor magnitude close to that of the P1 center. The broken N-C 
bond dilation was similar to that of P$1$ center in the diamonds.
\begin{table}[!t]
  \caption{Calculated hyperfine tensors (MHz) of  Be$_s$-N$_3$.
$\theta$ and $\phi$ are given as indicated in table~\ref{table:Be_iHFI}}
\label{table:Be_s-3N-HFI}
\center
  \begin{tabular}{lrlrlrl}
 \hline \hline
Species
    & \multicolumn{2}{c}{$A_1$}
    & \multicolumn{2}{c}{$A_2$}
    & \multicolumn{2}{c}{$A_3$}\\    \hline
\multicolumn{7}{c}{$C_{1}$, $S=1/2$}\\
Be           &$-0.2$   &$(50,-33)$ &$1.5$    &$(73,72)$  &$1.9$  &$(45,-180)$\\
N$_2$           &$87.2$   &$(73,-56)$ &$87.3$    &$(140,12)$  &$129.1$  &$(46,55)$\\
C           &$178.4$   &$(62,157)$ &$178.5$    &$(48,-83)$  &$362.6$  &$(45,56)$\\

\hline
 \hline \hline
  \end{tabular}
\end{table}

Finally, when the number of N atoms increased to four, the formation 
energy of Be$_s$-N$_4$ impurity reached $-4.64$\ev.
Be$_s$-N$_4$ structure exhibited $C_s$ symmetry [figure~\ref{fig:Bes-3N}(d)],
in contrast to a previous study~\cite{Yang-thesis}. The distances from Be to 
the three four-fold coordinated N atoms and three-fold coordinated N atom
 are  $1.65\angstrom$ and $1.63\angstrom$, respectively, which are 
 $7.1\%$ and
$5.8\%$ longer than the original C-C bond. The band-gap level of
 Be$_s$-N$_4$ is higher than that of Be$_s$-N$_3$, which is $1.16\ev$
 below the conduction band minimum. 
Midgap acceptors such as vacancies~\cite{dannefaer-DRM-10-2113} may ionize these defects 
without the need for illuminations. The effect of ionizing Be-N$_4$ is a
 reduction in the broken N-C dilation from $\%40$ in the neutral state to 
$\%29$ in the positive charge state compared to the C-C bond in a perfect diamond.
\begin{table}[!t]
	\caption{Calculated hyperfine tensors (MHz) of  (Be$_s$-N$_4$)$^{+1}$. 
		$\theta$ and $\phi$ are given as indicated in table~\ref{table:Be_iHFI}}
	\label{table:Be_s-4N-HFI}
	\center
	\begin{tabular}{lrlrlrl}
		\hline \hline
		Species
		& \multicolumn{2}{c}{$A_1$}
		& \multicolumn{2}{c}{$A_2$}
		& \multicolumn{2}{c}{$A_3$}\\    \hline
		\multicolumn{7}{c}{$C_{s}$, $S=1/2$}\\
		Be           &$-0.1$   &$(114.5,135)$ &$1.5$    &$(24.6,135)$  &$2$  &$(90-135)$\\
		N$_1$           &$86.1$   &$(143.5,135)$ &$86.2$    &$(90,45)$  &$127.2$  &$(53.5,135)$\\
		C            &$175.9$ &$(90,45)$ &$176$  &$(144,135)$ &$360$ &$(54.1,135)$\\
		
		\hline
		\hline 
	\end{tabular}
\end{table}
We also report the hyperfine interactions for Be, N and radical C atoms,
although to-date, there are no
experimental data with which to make a comparison. The unpaired 
electron density was localized in an N-C antibonding orbital similar
to that of the Be-N$_3$ complex. 

\section{Conclusions}\label{sec:discussion}

First-principles calculations were conducted to investigate the formation
 energy, solubility, and electronic properties of Be-doped and Be-N
 co-doped diamonds. The formation energy of substitutional Be defects
 ($4.62\ev$) is lower than that of interstitial Be defects ($10.28\ev$),
 suggesting that Be incorporation is more favorable as a substitutional
 dopant. However, the high formation energies indicate significant challenges
 in introducing Be dopants into the diamond structure. Co-doping with nitrogen
 was found to considerably reduce the formation energy, with Be-N complexes 
having low endothermic or exothermic formation energies depending on the
 number of N atoms. Among the interstitial Be configurations, the off-axis 
interstitial configuration with $C_s$ symmetry Be$_{i,oa1}$ was found to be
 the most stable. Substitutional Be (Be$_s$) introduces a double acceptor level
 $0.71 \ev$ above the valence band top. Be-N co-doping with increasing number 
of N atoms (1 to 4) progressively lowers the formation energy, with Be-N$_3$
 and Be-N$_4$ 
complexes exhibiting shallow donor levels at $1.0\ev$ and $1.16\ev$ below the 
conduction band minimum, respectively. Hyperfine tensors were calculated for
 various configurations to aid in future experimental identification of these
 defects.

  \bibliographystyle{cmpj}
  \bibliography{bibliography}
 
 \newpage
\ukrainianpart

\title
{Берилій та пов'язані з ним домішки в алмазі: дослідження методом функціоналу густини}
\author{К. М. Етмімі\refaddr{label1},
	М. А. Оджалах\refaddr{label2}, А. М. Аботрума\refaddr{label3}}
\addresses{
	\addr{label1} Фізичний факультет, факультет природничих наук, Університет Триполі, Триполі, Лівія
	\addr{label2} Фізичний факультет, освітній факультет, Університет Триполі, Триполі, Лівія
	\addr{label3} Вищий інститут передових професій (Інститут Аль-Шмох), Триполі, Лівія}

\makeukrtitle

\begin{abstract}
	\tolerance=3000%
	Для дослідження геометрії, електричних властивостей та надтонких структур різних конфігурацій алмазів, легованих берилієм, включаючи міжвузлові (Be$_i$), замісні (Be$_s$) та берилій-азотні (Be-N) комплекси, було застосовано першопринципне моделювання методом функціоналу густини. Включення Be в алмазну ґратку в якості замісної легуючої речовини є більш сприятливим, ніж міжвузловий легант, хоча обидва процеси є ендотермічними. Міжвузловий Be може потенційно  проявляти динамічне усереднення від площинної до осьової симетрії з енергією активації 0.1 еВ. Найстійкіша конфігурація Be$_s$ має симетрію $T_{d}$ зі спіновим станом $S=1$. Спільне легування азотом знижує енергію утворення комплексів Be$_s$-N, яка ще більше зменшується зі збільшенням кількості атомів азоту. Це пояснюється меншим ковалентним радіусом азоту у порівняні з вуглецем, що призводить до зменшення спотворення кристалічної ґратки. Спільне легування Be$_s$-N$_3$ та Be$_s$-N$_4$ вводить поверхневі донори, тоді як Be$_s$ демонструє напівпровідність $n$-типу, але глибокий донорний рівень робить його непрактичним для застосувань при кімнатній температурі. Отримані результати дають важливе розуміння поведінки берилію як легуючої домішки в алмазі та підкреслюють потенціал спільного легування берилієм і азотом для отримання алмазних напівпровідників $n$-типу.

	\keywords алмаз, берилій, азот, n-тип, p-тип, першопринципне моделювання
\end{abstract}

\end{document}